\documentclass[aps,preprint,showpacs,amsmath,amssymb]{revtex4-1}
\usepackage{graphicx}
\usepackage{dcolumn}
\usepackage{bm}
\usepackage{hyperref}

\begin{document}

\title{Genuine entanglement under squeezed generalized amplitude damping channels with memory}
\author{Mazhar Ali\footnote{Email: mazharaliawan@yahoo.com, mazhar.ali@iu.edu.sa}}
\address{Department of Electrical Engineering, Faculty of Engineering, Islamic University Madinah, 
107 Madinah, Saudi Arabia}

\begin{abstract}
We study genuine entanglement among $3$-qubits undergoing through a noisy process including dissipation, 
squeezing and decoherence. We obtain a general solution and analyze the asymptotic quantum states. It 
turns out that most of these asymptotic states can be genuinely entangled depending upon parameters of 
channel, memory parameter, and parameters of initial states. We study Greenberger-Horne-Zeilinger (GHZ) 
states and W states, mixed with white noise and determine the conditions for them to be genuinely entangled 
at infinity. We find that for these mixtures, it is possible to start with bi-separable state (with 
specific mixture of white noise as described below) and end up with genuine entangled states. However, 
the memory parameter $\mu$ must be very high. We find that in contrast to two-qubit 
case, all three qubit asymptotic states for $n \to \infty$ are not genuinely entangled.
\end{abstract}

\pacs{03.65.Yz, 03.65.Ud, 03.67.Mn}
\keywords{Genuine entanglement, squeezed thermal baths, multipartite states}

\maketitle

\section{Introduction}
\label{Sec:intro}

Quantum correlations among several particles, not only lead to counter-intuitive predictions but also 
have key role in future technologies \cite{Wilde-Book}. There is growing interest among researchers to 
study quantum correlations both in theory and experiments   
\cite{Horodecki-RMP81,gtreview,Erhard-NRP2,Friis-NRP1,Ali-QIP-2023}. 
Entanglement is a precious resource, which is important for several applications. One of the big 
challenge is to preserve quantum correlations in quantum states. Quantum states interact with their 
environments and it is well known that such interactions can degrade entanglement 
\cite{Yu1,Yu2,Yu3,Yu4}. Many authors have studied the effect of environments on entanglement for both 
bipartite and multipartite systems
\cite{Duer-PRL92,Hein-PRA71,Aolita-PRL100,Simon-PRA65,Borras-PRA79,Cavalcanti-PRL103,Band-PRA72,
Chaves-PRA82, Aolita-PRA82,Carvalho-PRL93,Lastra-PRA75,Guehne-PRA78,Lopez-PRL101,
Rau-EPL82,Ali-JPB42,Ali-JPB43,Ali-PRA81,Ali-JPB47,Weinstein-PRA85,Filippov-PRA88}.

Quantum channels describe the physical situations in which a given quantum state is transformed into another 
quantum state as a dynamical process. These processes are regarded as maps which preserve trace and are also 
complete positive, which means that a quantum state undergoing through a quantum channel is again a valid 
quantum state. 
One such channel is amplitude damping channel (ADC) which models spontaneous emission from excited atoms or 
energy dissipation at zero temperature \cite{Nielsen-Book}. This channel has been generalized to model 
dissipation at finite temperature \cite{Fujiwara-PRA70} and called generalized amplitude damping (GAD) 
channel. Another generalization of this model takes squeezing into account and the model is called squeezed 
generalized amplitude damping (SGAD) channel \cite{Srikanth-PRA77}. The effect on qubit-qubit 
entanglement for this channel was studied \cite{Banerjee-JPA40}, where it was found that this channel can 
not preserve entanglement. All these channels are considered as memoryless and effect of channel can be 
extend-able to $N$ qubits simply as $\Phi = \Phi_1^{\otimes N}$. There are situations where this 
simplification is not true and $\Phi \neq \Phi_1^{\otimes N}$ \cite{Banaszek-PRL92} and channel is said to 
have memory. The effect of memory on spin chains \cite{Plenio-PRL99}, entanglement-enhanced transmission of 
classical information in Pauli channels \cite{Daems-PRA76}, classical and quantum capacities of correlated 
amplitude damping channel \cite{Arrigo-PRA88} and others \cite{Caruso-RMP86,Guo-QIP16}, have been studied. 
Recently, qubit-qubit entanglement and quantum discord \cite{Ollivier-PRL88} under SGAD channel with memory 
has been studied \cite{Jeong-SR9}. 

An array of beam-splitters can be used to model a squeezed reservoir \cite{Kim-PRA52}. Laser-cooled trapped 
ions can also mimic the dynamics of an atom with a squeezed vacuum bath \cite{Poyatos-PRL77}. Another 
method to mimic coupling with squeezed bath via four level atoms driven by weak laser 
fields \cite{Lutkenhaus-PRA57}. Generalized amplitude damping channels are studied in experiments for 
decoherence and decay of atomic states. \cite{Myatt-Nature403, Turchette-PRA62}   

In this work, we study effect of memory on genuine entanglement for $3$-qubit systems, which to our 
knowledge has not been studied before. We consider three qubits sent by three consecutive use of quantum 
channel with memory. We determine the asymptotic states for a most general initial state of three 
qubits. We find that squeezing parameter $m$ is not present in the asymptotic states. This observation 
is in agreement with the fact that squeezed thermal bath can suppress quantum 
decoherence \cite{Banerjee-JPA40} but it is unable to preserve quantum 
entanglement \cite{Wilson-JMO50,Banerjee-AP325}. 
We then analyze these asymptotic states for various initial states and find that in most cases, they 
can be genuinely entangled depending on memory and parameter of initial states. We found that even if we 
start with bi-separable states then depending upon thermal parameter $n$ (must be not very large) and 
memory parameter $\mu$ (must be very large), asymptotic states 
can be genuinely entangled. Although the degree of genuine entanglement for these asymptotis states is 
quite small but it is an interesting feature of this dynamical process. 

This paper is organized as follows. In section \ref{Sec:Model}, we briefly discuss squeezed generalized 
amplitude damping channel for qubits and provide the general solution for an arbitrary state of $3$-qubits. 
We briefly review the concept of genuine entanglement in section \ref{Sec:GME} and present our main 
findings for two important families of quantum states. Finally, we conclude our work in 
section \ref{Sec:conc}.

\section{Squeezed generalized amplitude damping channel for qubits} 
\label{Sec:Model}

Quantum theory of damping takes a two-level atom (system) as harmonic oscillator interacting with 
reservoir (or bath) which can also be treated set of harmonic oscillators. The Hamiltonian in the 
interaction picture can be written as \cite{QO-Orszag, QO-Zubairy}
\begin{equation}
H = \hbar \, \sum_k \, g_k \, \big[ \, b_k^\dagger \, \sigma_- \, e^{-i(\omega -\nu_k)t} 
+ \sigma_+ \, b_k e^{i(\omega -\nu_k)t} \,\big]\,,  
\end{equation}
where $\sigma_- = |b\rangle\langle a|$ and $\sigma_+ = |a\rangle\langle b|$ are atomic lowering and 
raising operators. $b_k$ ($b_k^\dagger$) are bath annihilation (creation) operators for each mode. 
$\nu_k = c \, k$ are density distributed frequencies, $\omega$ is atomic transition frequency and 
$g_k$ are coupling constants. 
After standard quantum optical approximations and tracing over reservoir, we get 
a master equation for system only. Squeezed generalized amplitude damping channel is a noisy quantum 
channel in which a qubit interacts with a bath being initially in a squeezed thermal state with Markov 
and Born approximations. The master equation in interaction picture, is given as \cite{QO-Orszag,QO-Zubairy}
\begin{eqnarray}
\frac{d\, \rho}{dt} =& - \frac{\Omega \, (n+1)}{2} \bigg[\sigma_{+} 
\,\sigma_{-} \, \rho + \rho \, \sigma_{+} \, \sigma_{-} - 2\,\sigma_{-} \, \rho \, \sigma_{+} \bigg] 
- \frac{\Omega \, n}{2} \bigg[ \sigma_{-} \, \sigma_{+} \, \rho \nonumber \\& 
+ \rho \, \sigma_{-} \, \sigma_{+} - 2\, \sigma_{+} \, \rho \, \sigma_{-} \bigg] 
- \Omega \, m \, \bigg[\sigma_{+} \, \rho \, \sigma_{+} + \sigma_{-} \, \rho \, \sigma_{-} \bigg] \,. 
\label{Eq:ME}
\end{eqnarray}
$n$ is related with number of thermal photons and $m$ is the squeezing parameter. The complete positivity 
demands that for $\Omega \geq 0$, we must have $m^2 \leq n(n+1)$. We note that for $m = 0$, we have 
qubit interacting with thermal reservoirs, and for $n = m = 0$, we have (zero-temperature) amplitude damping 
process (vacuum reservoirs). 

We can extend this model for multi-qubits either as uncorrelated noise in which we have to sum these 
$3$ terms for each qubit separately and then solve the master equation in interaction picture by taking 
$\Omega_A = \Omega_B = \ldots \Omega_N = \Omega$, $n_A = n_B \ldots n_N = n$ and $m_A = m_B \ldots m_N = m$.
This process is the simplest model for memoryless quantum channel and the stochastic map 
$\Phi(\rho)$ can be extend-able to $\Phi^{\otimes N}(\rho)$ for $N$-systems or uses of quantum channel.
The result is that Kraus operators also have structure $K = K_A \otimes K_B \otimes \ldots K_N$. 
For a single qubit under such uncorrelated noise, the Kraus operators \cite{KO} can be 
written as
\begin{eqnarray}
K_1 &=& \left( \begin{array}{cc}
k_1 & 0 \\ 
0 & k_2   
\end{array} \right), \\
K_2 &=& \left( \begin{array}{cc}
0 &  k_3 \\ 
k_4 & 0  
\end{array} \right), \\
K_3 &=& \left( \begin{array}{cc}
\sqrt{e^{-\Omega \, t (n + 1/2)} \, cosh(\Omega \, t \, m)} &  0 \\ 
0 & \sqrt{e^{-\Omega \, t (n + 1/2)} \, cosh(\Omega \, t \, m)}  
\end{array} \right), \\
K_4 &=& \left( \begin{array}{cc}
0 & \sqrt{e^{-\Omega \, t (n + 1/2)} \, sinh(\Omega \, t \, m)} \\ 
\sqrt{e^{-\Omega \, t (n + 1/2)} \, sinh(\Omega \, t \, m)} & 0  
\end{array} \right), 
\end{eqnarray}\,
where 
\begin{eqnarray}
k_1 =& \sqrt{\frac{n}{2 n + 1} + \frac{n+1}{2 n +1} e^{-2 \, \Omega \, t (n + 1/2)} 
- e^{-\Omega \, t (n + 1/2)} \, cosh(\Omega \, t \, m)} \,, \\
k_2 =& \sqrt{\frac{n}{2 n + 1} e^{-2 \, \Omega \, t (n + 1/2)} + \frac{n+1}{2 n +1}  
- e^{-\Omega \, t (n + 1/2)} \, cosh(\Omega \, t \, m)} \,,\\ 
k_3 =& \sqrt{\frac{n}{2 n + 1} (1 - e^{-2 \, \Omega \, t (n + 1/2)}) - e^{-\Omega \, t (n + 1/2)} 
\, sinh(\Omega t m)} \, \\ 
k_4 =& \sqrt{\frac{n+1}{2 n + 1} (1- e^{-2 \, \Omega \, t (n + 1/2)}) - e^{-\Omega \, t (n + 1/2)} 
\, sinh(\Omega \, t \, m)} \,. 
\label{Eq:TE}
\end{eqnarray}
The Kraus operators satisfy the normalization condition $\sum_i \, K_i^\dagger(t) \, K_i(t) =  I$. 
For three qubits, there are $64$ such operators, that is, $M_1 = K_1^A K_1^B K_1^C$, 
$M_2 = K_1^A K_1^B K_2^C$, $\ldots$, $M_{64} = K_4^A K_4^B K_4^C$ with 
$\sum_{i=1}^{64} \, M_i^\dagger(t) \, M_i(t) =  I_8$. We have omitted the tensor product symbol between 
these operators. The stochastic map for SGAD with uncorrelated noise is 
\begin{eqnarray}
\Phi_u(\rho) = \rho_u (t) = \sum_{j=1}^{64} M_j \, \rho \, M_j^\dagger \,. 
\end{eqnarray}

For correlated noise $\Phi (\rho) \neq \Phi^{\otimes N} (\rho)$, and for three qubits, we have the correlated 
version of the master equation 
\begin{eqnarray}
\frac{d \rho}{dt} =& - \frac{\Omega \, (n+1)}{2} \bigg[\sigma_{+}^{\otimes 3} \,\sigma_{-}^{\otimes 3} 
\, \rho + \rho \, \sigma_{+}^{\otimes 3} \, \sigma_{-}^{\otimes 3} - 2\,\sigma_{-}^{\otimes 3} \, \rho \, 
\sigma_{+}^{\otimes 3} \bigg]  
- \frac{\Omega \, n}{2} \bigg[ \sigma_{-}^{\otimes 3} \, \sigma_{+}^{\otimes 3} \, \rho \nonumber \\&
+ \rho \, \sigma_{-}^{\otimes 3} \, \sigma_{+}^{\otimes 3} 
- 2\, \sigma_{+}^{\otimes 3} \, \rho \, \sigma_{-}^{\otimes 3} \bigg] 
- \Omega \, m \, \bigg[\sigma_{+}^{\otimes 3} \, \rho \, \sigma_{+}^{\otimes 3} + \sigma_{-}^{\otimes 3} 
\, \rho \, \sigma_{-}^{\otimes 3} \bigg] \,, 
\label{Eq:CN1}
\end{eqnarray}
where $\sigma_+^{\otimes 3} = \sigma_+ \otimes \sigma_+ \otimes \sigma_+$. The stochastic map for the 
process can be written either in terms of Kraus operators (which are not in tensor product format 
like uncorrelated case) or simply as solution of master equation (\ref{Eq:CN1}). In any case, it can be 
written as
\begin{eqnarray}
\Phi_c (\rho) = \rho_c (t) = \sum_{j} X_j \, \rho \, X_j^\dagger \,, 
\end{eqnarray}
where $X_j \neq X_A \otimes X_B \otimes \ldots X_N$. The density matrix for an arbitrary state of $3$ qubits 
has a simple solution in this case
\begin{eqnarray}
\rho_c (t) =  \left( \begin{array}{cccccccc}
\rho_{11}(t) &  \rho_{12}(t) & \rho_{13}(t) & \rho_{14}(t) & \rho_{15}(t) & \rho_{16}(t) & \rho_{17}(t) 
& \rho_{18}(t) \\ 
\rho_{21}(t) &  \rho_{22} & \rho_{23} & \rho_{24} & \rho_{25} & \rho_{26} & \rho_{27} & \rho_{28}(t) \\
\rho_{31}(t) &  \rho_{32} & \rho_{33} & \rho_{34} & \rho_{35} & \rho_{36} & \rho_{37} & \rho_{38}(t) \\
\rho_{41}(t) &  \rho_{42} & \rho_{43} & \rho_{44} & \rho_{45} & \rho_{46} & \rho_{47} & \rho_{48}(t) \\
\rho_{51}(t) &  \rho_{52} & \rho_{53} & \rho_{54} & \rho_{55} & \rho_{56} & \rho_{57} & \rho_{58}(t) \\
\rho_{61}(t) &  \rho_{62} & \rho_{63} & \rho_{64} & \rho_{65} & \rho_{66} & \rho_{67} & \rho_{68}(t) \\
\rho_{71}(t) &  \rho_{72} & \rho_{73} & \rho_{74} & \rho_{75} & \rho_{76} & \rho_{77} & \rho_{78}(t) \\
\rho_{81}(t) &  \rho_{82}(t) & \rho_{83}(t) & \rho_{84}(t) & \rho_{85}(t) & \rho_{86}(t) & \rho_{87}(t) 
& \rho_{88}(t)
\end{array} \right)\,, \label{Eq:CN}
\end{eqnarray}
where 
\begin{eqnarray}
\rho_{1s} (t) &=& \rho_{1s}(0) \, e^{-\frac{\Omega \, t (n+1)}{2}} \nonumber \\
\rho_{s8} (t) &=& \rho_{s8}(0) \, e^{-\frac{\Omega \, t \, n}{2}} \nonumber \\
\rho_{11}(t) &=& \frac{n (\rho_{11} + \rho_{88})}{2n+1} + \frac{ \rho_{11} + n (\rho_{11} - \rho_{88} ) }{2n+1} 
e^{-(2n+1) \Omega t } \nonumber \\
\rho_{18}(t) &=& \frac{1}{2} \big[\rho_{18} + \rho_{81} + e^{2 m \Omega t} (\rho_{18} - \rho_{81})\big] 
e^{-(n + m + 1/2) \Omega t} \nonumber \\
\rho_{88}(t) &=& \frac{(1-e^{-(2n+1) \Omega \, t}) (1+n) \rho_{11}}{2n+1} + \frac{1+n (1+e^{-(2n+1) \Omega \, t}) 
\, \rho_{88}}{2n+1}\,, 
\end{eqnarray}
with $s = 2,3,\ldots,7$. Finally, the stochastic map for three qubits sent by three consecutive use of 
channel with memory can be written as
\begin{equation}
\rho(t) = \mu \, \Phi_c(\rho)(t) + (1 - \mu) \, \Phi_u(\rho)(t)\,,
\end{equation}
where $0\leq \mu \leq 1$ is degree of channel memory, which means that the noise is correlated with 
probability $\mu$. We have the most general solution for the system and we can study the asymptotic states 
by taking $t \to \infty$. Examples below refer to such states as $\rho(\infty)$. Although, individual matrix 
elements have quite lengthy expressions, however, it is possible to study evolution of entanglement 
numerically using MATLAB as described in section below. 

\section{Genuine entanglement under SGAD channels}
\label{Sec:GME}

It is appropriate that we first briefly review the ideas about genuine entanglement. Let us consider 
$3$-qubits to discuss the ideas with this understanding that same arguments are equally valid for 
multipartite systems where each subsystem is a qudit. For pure states, we say that a state is fully 
separable if $|\psi^{ABC}\rangle = |\psi^A\rangle \otimes |\psi^B\rangle \otimes |\psi^C\rangle$. 
For mixed states, a state is fully separable if 
$\rho = \sum_j p_j \rho_j^A \otimes \rho_j^B \otimes \rho_j^C$. Bi-separable states can be written as  
$\rho_{A|BC}^{sep} = \sum_j \, q_j \, |\phi_A^j \rangle\langle \phi_A^j| \otimes |\phi_{BC}^j \rangle\langle \phi_{BC}^j|$, 
with similar construction for $\rho_{B|AC}^{sep}$ and $\rho_{C|AB}^{sep}$. Their mixtures are also 
bi-separable $ \rho^{bs} = p_1 \, \rho_{A|BC}^{sep} + p_2 \, \rho_{B|AC}^{sep} + p_3 \, \rho_{C|AB}^{sep}$.
A multipartite state is said to be genuinely entangled if it is not fully separable and not bi-separable. 
We should keep in mind that there are quantum states which have negative partial transpose (NPT) under each 
partition but they are bi-separable \cite{gtreview}. 

Genuine entanglement can be detected and quantified via positive partial transpose mixtures (PPT mixtures)
\cite{Bastian-PRL106,Novo-PRA88,Hofmann-JPA47}. PPT mixtures are characterized with semidefinite 
programming (SDP) and this approach can be used to quantify genuine entanglement \cite{Bastian-PRL106}. 
For bipartite systems, this procedure is equivalent to {\it negativity} \cite{Vidal-PRA65}, so we can 
call it as genuine negativity. For multi-qubits, it is bounded by $E(\rho) \leq 1$ with upper bound for 
certain pure states \cite{Jungnitsch-PRA84}. For mixed states it is always less than $1$. If this measure is 
positive then state is guaranteed to be genuine entangled, otherwise we are not sure unless some other 
procedure indicates its entanglement properties.  

It is well known that for $3$-qubits there are two inequivalent families of quantum states, namely, 
$GHZ$ states and $W$ states, given as  
\begin{eqnarray}
|GHZ_1 \rangle &=& \frac{1}{\sqrt{2}}(|000\rangle + |111\rangle), \nonumber \\
|W\rangle &=& \frac{1}{\sqrt{3}}(|001\rangle + |010\rangle + |100\rangle).
\label{Eq:GHZ3Qb1}
\end{eqnarray}
Other equivalent GHZ states are $|GHZ_2 \rangle = 1/\sqrt{2}(|001\rangle + |110\rangle)$, 
$|GHZ_3 \rangle = 1/\sqrt{2}(|010\rangle + |101\rangle)$, and 
$|GHZ_4 \rangle = 1/\sqrt{2}(|011\rangle + |100\rangle)$. $W$ state equivalent is 
$|\tilde{W}\rangle = 1/\sqrt{3}(|011\rangle + |101\rangle + |110\rangle)$. $GHZ$ states 
have maximum value of genuine negativity, that is, $E(|GHZ_j\rangle\langle GHZ_j|) = 1$, whereas for 
the $W$ state, $E(|W\rangle\langle W|) = E(|\tilde{W}\rangle\langle \tilde{W}|) \approx 0.886$ 
\cite{Bastian-PRL106}. 

As a first example, we consider $GHZ_1$ state mixed with white noise
\begin{equation}
\rho_1 = \alpha \, |GHZ_1\rangle\langle GHZ_1| + \frac{1 - \alpha}{8} \, I_8 \,,
\end{equation}
where $0 \leq \alpha \leq 1$. It is known that these states are genuinely entangled for 
$0.429 \leq \alpha \leq 1$ \cite{Bastian-PRL106}. An interesting property of SGAD channel with memory is
that for these states, the only non-zero matrix elements are only on the main diagonal and main off-diagonal 
of the density matrix whereas all other elements remain zero ($X$ structure) \cite{Rau-JPA42}. 
A result on the detection of genuine entanglement states that the inequality 
\begin{equation}
 |\rho_{18}| \leq \sqrt{\rho_{22} \, \rho_{77}} + \sqrt{\rho_{33} \, \rho_{66}} + \sqrt{\rho_{44} \, \rho_{55}}
\end{equation}
is satisfied by bi-separable states and the violation implies genuine entanglement \cite{Otfried-NJP12}. 
This criterion is a necessary and sufficient condition for GHZ-diagonal states \cite{Otfried-NJP12}. 
Application of this result to $\rho_1(\infty)$ gives us condition that 
\begin{eqnarray}
3 \sqrt{\bigg( \, \frac{n^2 (1+n)(1-\mu)}{(1+2n)^3} + \mu (1-\alpha)/8 \bigg) 
\bigg( \, \frac{n(1+n)^2(1-\mu)}{(1+2n)^3} + \mu (1-\alpha)/8 \bigg)} \nonumber \\ 
\geq \frac{(n(1+n))^{3/2} \, \alpha \, (1-\mu)}{(1+2n)^3} \, .   
\end{eqnarray}
This condition is always satisfied for $ n \to \infty$ as it simplifies to $\mu \leq (3 - \alpha)/2\alpha$. 
Hence all asymptotic states with $n \to \infty$ are 
not genuinely entangled. However, it is possible that condition is violated for certain values of 
$\alpha$, $n$ and $\mu$. Therefore, it is possible to have asymptotic states which are genuinely entangled. 

\begin{figure}[t!]
\scalebox{2.30}{\includegraphics[width=1.99in]{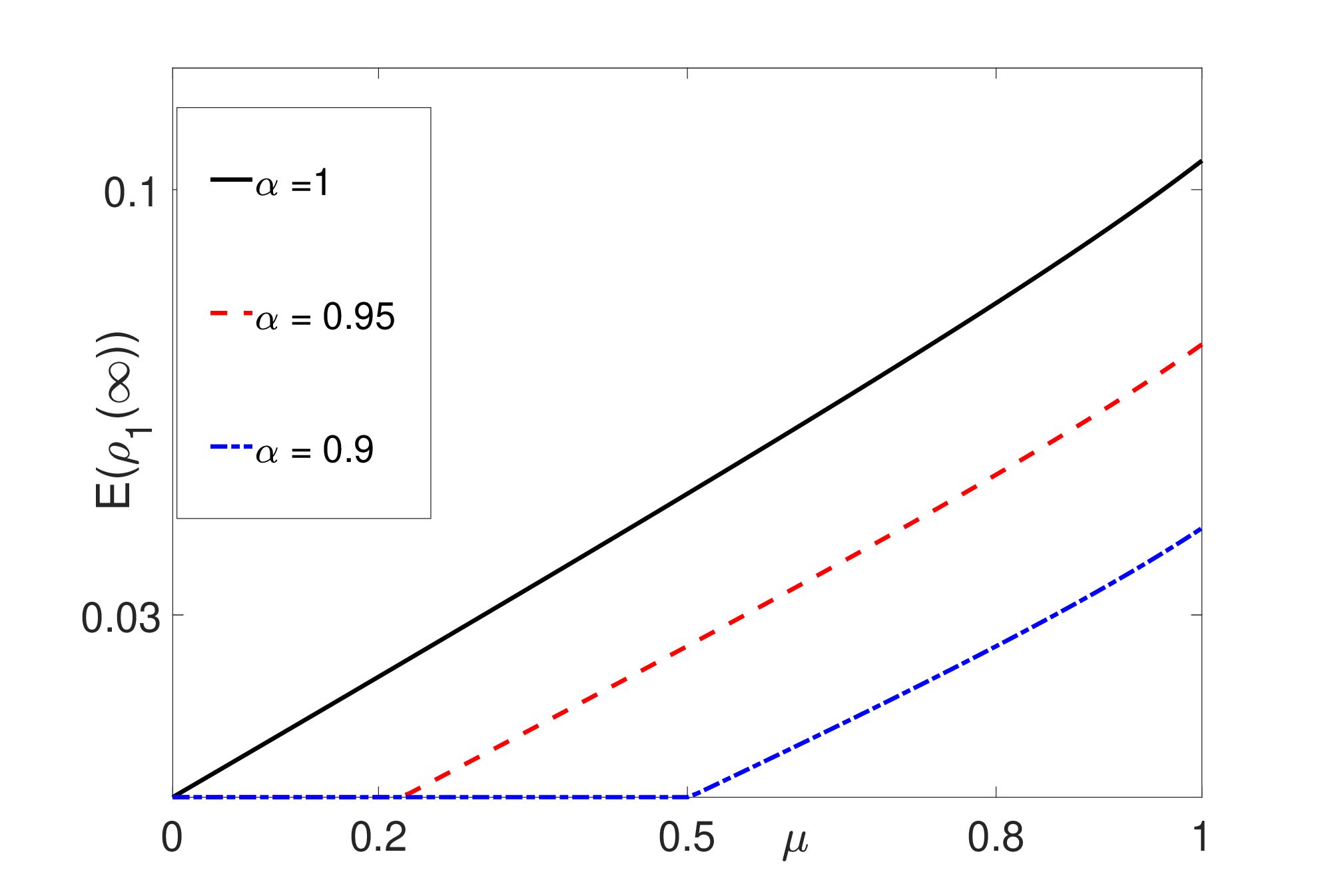}}
\centering
\caption{Genuine entanglement for states at infinity is plotted against memory parameter $\mu$ and $n = 1$. 
As we see that for $\alpha = 1$ asymptotic states are genuine entangled for $\mu > 0$. If $\alpha < 1$, 
then some states are genuinely entangled if we increase parameter $\mu$.}
\label{Fig:ex1}
\end{figure}
In Figure (\ref{Fig:ex1}), we plot genuine entanglement $E(\rho)$ for asymptotic states with $n = 1$. 
We observe that as long as memory $\mu > 0$ the state $\rho_1 (\infty)$ with $\alpha = 1$ is always genuinely 
entangled. For mixture with white noise, if we take $\alpha = 0.95$ (dashed line) or $\alpha = 0.9$ 
(dotted-dashed line), the asymptotic states are genuinely entangled for larger values of memory $\mu$. 
We also observe that most of genuine entanglement is lost as asymptotic states have much less degree of 
genuine entanglement. In this case, our initial states are genuinely entangled and asymptotic states may 
be genuinely entangled as seen in Figure (\ref{Fig:ex1}).

\begin{figure}[t!]
\scalebox{2.30}{\includegraphics[width=1.99in]{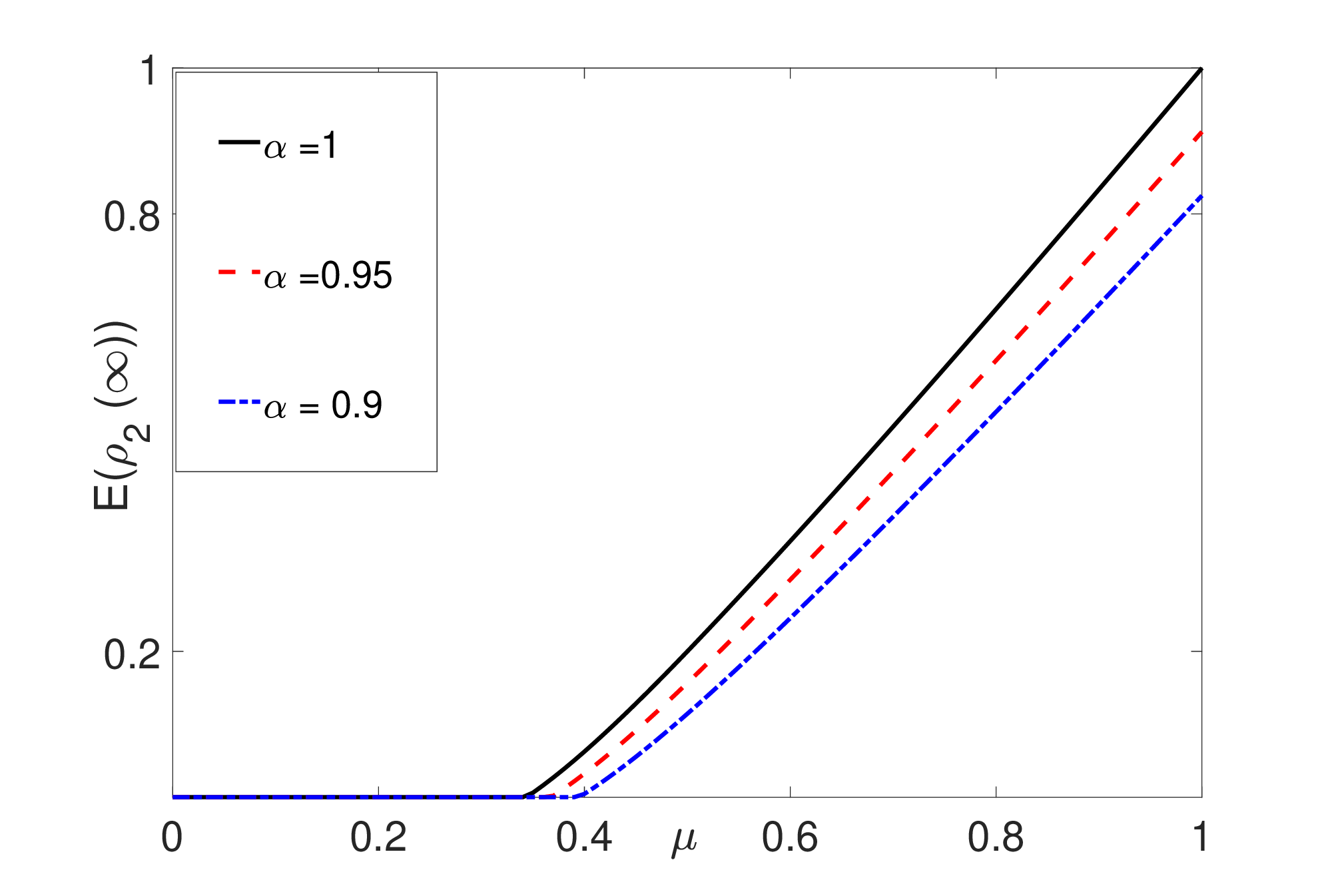}}
\centering
\caption{Multipartite entanglement is plotted for asymptotic states against memory $\mu$. We took $n = 1$ 
for this plot.}
\label{Fig:ex2}
\end{figure}

In contrast, if we take initial state $|GHZ_2\rangle$ mixed with white noise, we expect different results. 
As we saw in Eq.(\ref{Eq:CN}) $|GHZ_2\rangle$ lives in decoherence free space, therefore its entanglement 
is not changing and for $\mu = 1$ and $\alpha = 1$, we must have the same state at infinity. 
For $\alpha < 1$ and $\mu < 1$, we have to analyze the asymptotic states. 
In Figure~(\ref{Fig:ex2}), genuine entanglement is plotted for asymptotic states with $n = 1$ for three 
choices of parameter $\alpha$. We can see that for $\mu = 1$, and $\alpha = 1$, the states are stationary and 
genuine entanglement is fixed at maximum value of $1$. However, as memory $\mu$ and $\alpha$ are decreased 
then effects of uncorrelated noise degrades the genuine entanglement until for a specific value of memory, 
asymptotic states are no longer genuinely entangled.  

\begin{figure}[t!]
\scalebox{2.30}{\includegraphics[width=1.99in]{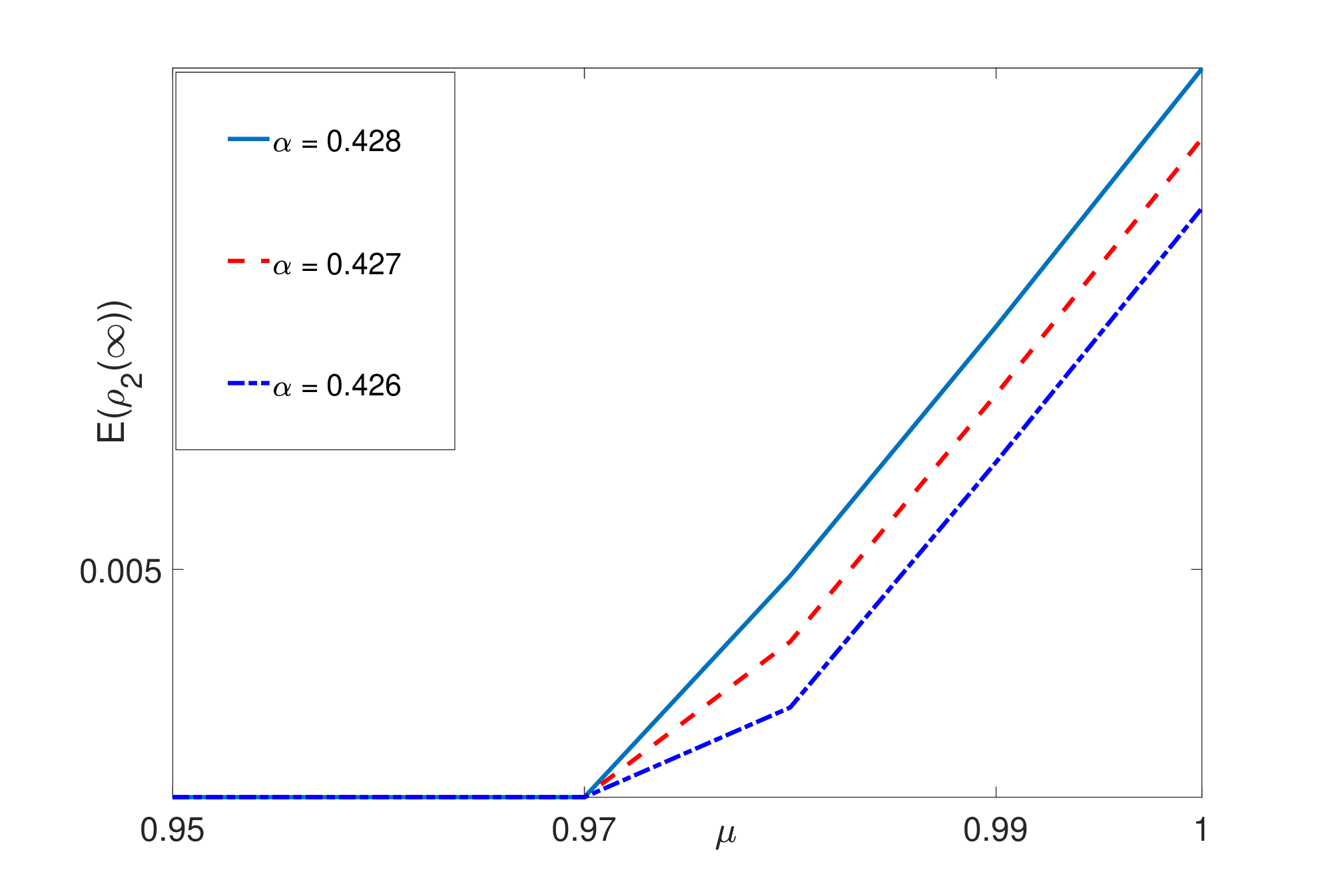}}
\centering
\caption{Genuine entanglement is plotted against memory $\mu$. This plot is for $n = 0.1$. We observe 
that all $3$ initially bi-separable states may become genuine entangled as a result of dynamical process.}
\label{Fig:ex2a}
\end{figure}

It turns out that condition for bi-separable states for $n \to \infty$ gives us 
$\mu \leq \frac{6 - \alpha}{13 \, \alpha}$, which is not true for whole range of $\alpha$. It is violated 
for $\alpha < 0.429$ and the initial states are genuinely entangled for 
$ 0.429 \leq \alpha \leq 1$. As this result is for very large $n$, therefore, we expect that it is possible 
to start with bi-separable state and end up with genuine entangled state, provided that memory $\mu$ is 
quite high and $n$ is small. In Figure~(\ref{Fig:ex2a}), we observe that all $3$ bi-separable states can end 
up with genuine entangled states for $\mu \geq 0.97$.

Let us now take another type of genuine entangled state, namely, $W$ state mixed with white noise, as
\begin{equation}
\rho_W = (1-\beta) \, |W \rangle\langle W| + \frac{\beta}{8} \, I_8 \,, 
\end{equation}
where $0 \leq \beta \leq 1$. We want to mention here that $1-\beta = \alpha$ has the same meanings and 
we could have used $\alpha$ instead of $\beta$. These states are genuinely entangled for 
$0 \leq \beta \leq 0.521$ \cite{Bastian-PRL106}. 

\begin{figure}[t!]
\scalebox{2.30}{\includegraphics[width=1.99in]{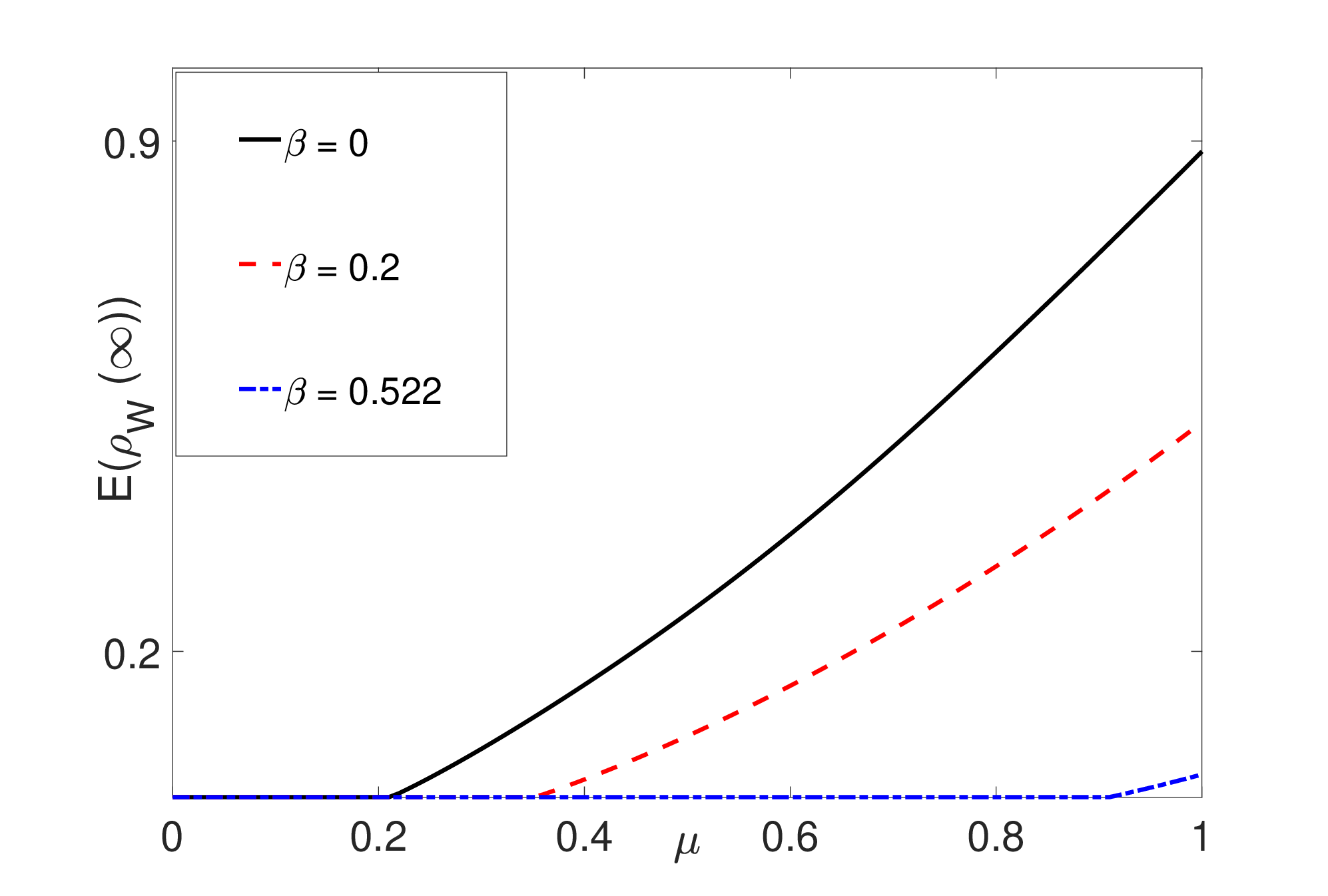}}
\centering
\caption{Genuine entanglement plotted against memory $\mu$ with $n = 1$ for initially $W$ states mixed 
with white noise.}
\label{Fig:ex3}
\end{figure}

Figure (\ref{Fig:ex3}) depicts genuine entanglement for asymptotic states with initial states as genuine 
entangled for $\beta = 0,\, 0.2$ and bi-separable for $\beta = 0.522$. We observe that even though initial 
state is bi-separable (for $\beta = 0.522$) but asymptotic state is genuinely entangled. We observe that 
all entangled asymptotic states have very small degree of genuine entanglement, nevertheless, it is an 
interesting feature of this channel that it can convert bisepaeable states into genuine entangled states. 

\section{Conclusions and Discussions}
\label{Sec:conc}

We have studied genuine entanglement under SGAD channel with memory for three qubits. We have obtained the 
most general solution for the system. Based on this solution, we are able to analyze the asymptotic states. 
We have found that in the case of $|GHZ_1\rangle$ states mixed with white noise and for $n \to \infty$, all 
asymptotic states are not genuinely entangled. However, for small values of $n$ and depending on memory $\mu$ 
and initial states, it is possible to have genuine entangled asymptotic states. We have also 
taken initial biseparable states and have found that under certain circumstances the asymptotic states 
may be genuine entangled. We found this possibility not only in $GHZ$ states but also for $W$ states mixed 
with white noise. It is an interesting feature of this dynamical process that it may bring biseparable states 
to genuine entangled ones. However for this phenomenon to occur, we must have very high degree of memory 
$\mu$ in the channel. 
We also observed that squeezing parameter $m$ is absent only among asymptotic states. It is well known 
that expectation values of certain physical quantities at infinity are function of average thermal 
photons $\langle n(\omega)\rangle$ \cite{QO-Orszag}. This has a simple interpretation. After a long time, 
the oscillator in contact with a heat bath gets thermalized, with the same average photon number as the 
thermal average at oscillator's frequency. The difference between two-qubits under SGAD and three qubits 
under similar noise can be summarized as follows. It was found that for two-qubits singlet state with 
correlated noise, squeezing parameter $m$ does not effect the dynamics of entanglement and quantum 
discord \cite{Jeong-SR9}. In addition some of the two-qubit asymptotic states are also not dependent on 
squeezing parameter $m$. Third, it is possible that with singlet states as initial states, the asymptotic 
states with $n \to \infty$ can be entangled even if initial states are separable. For three qubits, 
we get similar results with some differences. The first main difference is the fact that we can only get 
genuine entangled states at infinity if $n$ is not too large, because we observed that all asymptotic states 
with $n \to \infty$ are not genuinely entangled. Similar to two qubits, for correlated noise $\mu = 1$, 
GHZ state living in decoherence free space is invariant under dynamics. For three qubits, we have 
two types of inequivalent genuine entangled states, whereas for two qubits, states are either entangled 
or separable.

\acknowledgements{The author is grateful to reviewers for their constructive and meticulous comments which 
brought much clarity in the manuscript.}


\begin{thebibliography}{99}
\bibitem{KO} We figured out that Kraus operators provided in Ref.\cite{Jeong-SR9} have typos and also 
their number is 6. We provide here only 4 operators satisfying the standard restrictions on them.
\end{thebibliography}
\end{document}